\documentclass[%
 aip,
 amsmath,amssymb,
 reprint,%
]{revtex4-1}

\usepackage{graphicx}
\usepackage{dcolumn}
\usepackage{bm}

\usepackage[utf8]{inputenc}
\usepackage[T1]{fontenc}
\usepackage{mathptmx}
\usepackage{etoolbox}

\makeatletter
\def\@email#1#2{%
 \endgroup
 \patchcmd{\titleblock@produce}
  {\frontmatter@RRAPformat}
  {\frontmatter@RRAPformat{\produce@RRAP{*#1\href{mailto:#2}{#2}}}\frontmatter@RRAPformat}
  {}{}
}%
\makeatother
\begin{document}

\preprint{AIP/123-QED}

\title[Competing random searchers under restarts]{Competing random searchers under restarts}
\author{R. K. Singh}
 \affiliation{Department of Biomedical Engineering, Ben-Gurion University
 of the Negev, Be'er Sheva 85105, Israel}
\affiliation{Department of Physics, Bar-Ilan University, Ramat-Gan 5290002,
Israel}
 \email{rksinghmp@gmail.com}
 
\author{R. Metzler}%
\affiliation{ 
Institute of Physics and Astronomy, University of Potsdam, D-14476 Potsdam-Golm, Germany} \affiliation{ 
Asia Pacific Center for Theoretical Physics, Pohang 37673, Republic of Korea}%

\author{T. Sandev} 
\affiliation{Research Center for Computer Science and Information Technologies, Macedonian Academy of Sciences and Arts, Bul. Krste Misirkov 2, 1000 Skopje, Macedonia} \affiliation{Institute of Physics, Faculty of Natural Sciences and Mathematics, Ss. Cyril and Methodius University, Arhimedova 3, 1000 Skopje, Macedonia}
\affiliation{Department of Physics, Korea University, Seoul 02841, Republic of Korea} 
%

\date{\today}

\begin{abstract}
We study independent searchers competing for a target under restarts and
find that introduction of restarts tends to enhance the search efficiency
of an already efficient searcher. As a result, the difference between the
search probabilities of the individual searchers increases when the system
is subject to restarts. This result holds true independent of the identity
of individual searchers or the specific details of the distribution of
restart times. However, when only one of a pair of searchers is subject to restarts
while the other evolves in an unperturbed manner, a concept termed as
\textit{subsystem restarts}, we find that the search probability exhibits
a nonmonotonic dependence on the restart rate. We also study the mean search
time for a pair of run and tumble and Brownian searchers when only the
run and tumble particle is subject to restarts. We find that, analogous to
restarting the whole system, the mean search time exhibits a nonmonotonic
dependence on restart rates.
\end{abstract}

\maketitle

\begin{quotation}
Random walks searching for a target under restarts have been extensively
studied. However, the general case of multiple walkers searching for multiple
targets has received rather limited attention. We study independent searchers
competing for a target under restarts with primary focus on search probabilities,
thus providing a complementary view to an analysis based on first passage times.
One of the main findings of this work is the existence of an extremum of search
probabilities when the system is subject to subsystem restarts.
\end{quotation}

\section{\label{sec1}Introduction}
Stochastic restarts have emerged as one of the most widely investigated
concepts of nonequilibrium statistical physics during the last decade
\cite{gupta2022stochastic}. The fundamental essence of restarting a random
system is that its dynamical state is renewed at certain intervals of time
which can either be a random variable or a fixed constant, thus modifying the
dynamical properties of the system in a nontrivial manner. For example,
restarting a Brownian motion at fixed rates leads to a nonequilibrium steady state
\cite{singh2022general},
additionally rendering the mean search time finite \cite{evans2011diffusion,singh2021backbone}.
Restarts have been employed as means of speeding up algorithms in computer science
\cite{luby1993optimal} and have emerged as a way to expedite target searches
in physics and chemistry \cite{reuveni2016optimal,pal2017first}. A common
feature of many of these studies is a single searcher searching
for a single target \cite{evans2011JPA,eliazar2020mean,eliazar2021mean,pal2016diffusion,
bhat2016stochastic,
evans2018run,masoliver2019telegraphic,nagar2016diffusion,domazetoski2020stochastic,
evans2014diffusion,ahmad2019first,ahmad2022first}. In other words, the search process
subject to restarts stops the moment the searcher detects the target. However, in most
cases of practical interest either there exist more than one target \cite{pal2019first,
chechkin2018random} or multiple searchers are searching for one or more targets
\cite{evans2022exactly,belan2018restart,singh2022capture,singh2023bernoulli,vilk2022fluctuations,
nagar2023stochastic}. In the latter case of multiple searchers competing for
a single target a naturally arising question is about the search efficiency of
a given searcher. In other words, if there is a pair of searchers competing for
a target, then which of the two possesses a higher probability to reach the target?
While the answer is trivial for identical searchers searching in identical
conditions, the situation in which either the conditions are non-identical, say
different initial conditions, or moreover, the searchers themselves are distinct,
is far from trivial. Furthermore, an assessment of the search probabilities is complementary
to an estimate of (un)conditional first passage times; and even though the latter has
been extensively studied under restarts \cite{evans2019stochastic}, the former
remains relatively less explored \cite{belan2018restart,singh2023bernoulli}.
In this paper we will study in detail the search probabilities for a pair of
searchers competing for a target.

Consider a pair of searchers $S_1$ and $S_2$ moving independently in one dimension
and searching for a target located at the origin. The initial location of
the two searchers is $x_i(0) = c_i,~i \in \{1,2\}$. At this point, the two searchers $S_1$
and $S_2$ can either be identical (like a pair of Brownian searchers) or distinct
(a Brownian searcher competing with a run and tumble particle (RTP)). The latter case
of distinct searchers is particularly relevant from the point of view of biological
searches wherein it is not uncommon to observe L\'{e}vy searchers
competing with Brownian particles \cite{bartumeus2002optimizing,james2008optimizing,palyulin2014levy,palyulin2019first} or combined L\'evy-Browninan search \cite{palyulin2016search}.
Furthermore, any search strategy which involves revisiting previously
visited locations is likely to be less efficient compared
to a simple ballistic search \cite{raposo2003dynamical}. Does it mean
that the scenario in which a L\'{e}vy searcher and a Brownian searcher are
searching for a target has a clear winner? While the previous studies on
optimizing encounter rates in biological interactions
answer this question in terms of efficiency of different searchers  
\cite{bartumeus2002optimizing,james2008optimizing,palyulin2014levy},
the effect of restarts remains unexplored. In light of the extensive
amount of research on the topic, a study on the search properties of competing
searchers poses a timely question. Now it is intuitively expected that
the closer the searcher starts to the target, the higher is the probability of a successful
detection. In other words, if $c_1 < c_2$ then $p_1 > p_2$ where $p_i,~i \in \{1,2\}$,
is the search probability of the searcher $S_i$. In what follows, we will show this
to be true for a pair of independent Brownian searchers and study the properties
of search probabilities $p_1$ and $p_2$ in the presence of restarts.
The only constraint imposed on the search probabilities follows from the
dimensionality of the search space leading to $p_1 + p_2 = 1$.

\section{\label{sec2}Competing searchers searching for a target}
Consider a system of two otherwise non-interacting Brownian particles
$B_1$ and $B_2$ searching for a target located at the origin. Let us further
assume without any loss of generality that both particles have identical
diffusion coefficients $D$. As a result, the equation of motion for the joint probability density function (PDF) $p \equiv p(x_1,x_2,t)$ is
\begin{align}
\label{dyn}
\frac{\partial}{\partial t} p(x_1,x_2,t) = D \frac{\partial^2}{\partial x_1^2} p(x_1,x_2,t)
+ D \frac{\partial^2}{\partial x_2^2} p(x_1,x_2,t).
\end{align}
As the two searchers are searching for a target located at the origin,
we have the following boundary conditions: $p(0,x_2,t) = 0$ ($B_1$ reaches
the target) and $p(x_1,0,t) = 0$ ($B_2$ reaches the target).
In addition, the initial conditions are chosen as
$p(x_1,x_2,0) = \delta(x_1-c_1)\delta(x_2-c_2)$ with $0 < c_1 < c_2$.
Eq.~(\ref{dyn}) can be solved exactly using the method of images \cite{fisher1984walks,
redner2001guide,metzler2014first} and its solution reads
\begin{widetext}
\begin{align}
\label{sol}
p(x_1,x_2,t) = f(x_1,x_2,t) - f(-x_1,x_2,t) + f(-x_1,-x_2,t) - f(x_1,-x_2,t),
\end{align}
\end{widetext}
where $f(x_1,x_2,t) = \frac{1}{4\pi Dt}\exp\Big[-\frac{(x_1-c_1)^2+(x_2-c_2)^2}
{4Dt}\Big]$ is the joint PDF of a couple of independent and identical
Brownian particles moving on the real line $(-\infty,\infty)$. It is evident
from Eq.~(\ref{sol}) that the solution respects both the initial and boundary
conditions. Furthermore, even though the two searchers do not interact
with each other (crossing of trajectories is allowed), the search process
stops at the moment when the target is detected by any of the two walkers.
In other words, the dynamics of $B_1$ and $B_2$ is independent only as long
as the search is ongoing.
In order to estimate the search probabilities, we study the survival probability that
none of the searchers crosses the origin up to time $t$, which is given by
\begin{align}
q_{B_1B_2}(t) = \int^\infty_0 dx_1 \int^\infty_0 dx_2 p(x_1,x_2,t).
\end{align}
As a result, the unconditional first passage time density (FPTD):
$F_{B_1B_2}(t) = -\frac{d}{dt}q_{B_1B_2}(t)$ reads
\begin{widetext}
\begin{align}
\label{fpt}
F_{B_1B_2}(t) = \frac{c_1}{\sqrt{4\pi Dt^3}}\exp\Big(-\frac{c^2_1}{4Dt}\Big)\times
\textrm{erf}\Big(\frac{c_2}{\sqrt{4Dt}}\Big)
+ \textrm{erf}\Big(\frac{c_1}{\sqrt{4Dt}}\Big) \times \frac{c_2}{\sqrt{4\pi Dt^3}}
\exp\Big(-\frac{c^2_2}{4Dt}\Big).
\end{align}
\end{widetext}
This is a very interesting result and can be intuitively arrived at by using
the additive and multiplicative rules of probability for a pair of independent
events. The first term in Eq.~(\ref{fpt}) is the product of two events:
first is the target being detected by $B_1$ while $B_2$ is still
meandering on the semi-infinite line. The second term of Eq.~(\ref{fpt})
has the same interpretation with $B_1$ and $B_2$ interchanged.
Now, expanding the error and exponential functions for small arguments we
can arrive at the long time behavior of the FPTD, and find (to leading order):
\begin{align}
\label{lt}
F_{B_1B_2}(t) \stackrel{t \to \infty}{\sim} \frac{2}{\pi t^2}.
\end{align}
In other words, even though the two searchers are non-interacting, they do
feel each others' presence, as for a single searcher
$F_{B_1B_2}(t) \stackrel{t \to \infty}{\simeq}
\frac{1}{t^{3/2}}$, a slower decay compared to the
FPTD for a pair of Brownian searchers (see Eq.~(\ref{lt})). However,
the relatively faster decay of the unconditional FPTD
$F_{B_1B_2}(t)$ does not prevent the divergence of moments, as
\begin{align}
\label{mfpt}
\langle t \rangle = \int^\infty_0 dt~t F_{B_1B_2}(t)
\stackrel{\text{large}~t}{\simeq} \ln t|^\infty.
\end{align}
In other words, the mean time of search taken by the pair of independent Brownian
searchers to locate the target is not finite. Now
from Eq.~(\ref{fpt}), we can easily write the FPTD
conditioned on the information as to which searcher ($B_1$
or $B_2$) completes the search. Using $F_{B_1B_2}(t) = F_{B_1}(t) + F_{B_2}(t)$
we find
\begin{subequations}
\label{cond}
\begin{align}
F_{B_1}(t) &= \frac{c_1}{\sqrt{4\pi Dt^3}}e^{-\frac{c^2_1}{4Dt}}
\textrm{erf}\Big(\frac{c_2}{\sqrt{4Dt}}\Big),~\text{target located by $B_1$},\\
F_{B_2}(t) &= \textrm{erf}\Big(\frac{c_1}{\sqrt{4Dt}}\Big)\frac{c_2}{\sqrt{4\pi Dt^3}}
e^{-\frac{c^2_2}{4Dt}},~\text{target located by $B_2$}.
\end{align}
\end{subequations}
It is evident from the above equations that for $c_1 = c_2$ both $B_1$
and $B_2$ have equal search probabilities.
Using $c_1 = 1,~c_2 = 2$ and $D = 1$ we find that $\int^\infty_0 dt~F_{B_1}(t)
\approx 0.7$ and $\int^\infty_0 dt~F_{B_2}(t) \approx 0.3$, just what we would
intuitively expect, that is, a searcher starting close to the target would have a
higher likelihood of locating the target.

Interestingly, the analysis gets fairly involved when the competing searchers
are non-identical, a scenario important from the point of view of biological
searches \cite{bartumeus2002optimizing,james2008optimizing}. Notwithstanding,
we address this question by considering a pair of RTP $R_1$ and a Brownian particle
$B_1$ searching for a target.
The equations of motion describing the dynamics of the two particles is
\begin{subequations}
\label{langRB}
\begin{align}
\frac{dx_1}{dt} &= v\sigma(t) + \eta_1(t),\\
\frac{dx_2}{dt} &= \eta_2(t),
\end{align}
\end{subequations}
where $v$ is the velocity of $R_1$ and $\sigma(t)$ switches between $\pm 1$
at a Poisson rate $\gamma$ with correlation $\langle \sigma(t) \sigma(t') \rangle
= \exp(-2\gamma|t-t'|)$. The two independent noise sources $\eta_1(t)$ and $\eta_2(t)$
are Gaussian with mean zero and delta-correlation: $\langle \eta_i(t) \eta_j(t') \rangle
= \sqrt{2D}\delta_{ij}\delta(t-t'),~i,j\in\{1,2\}$, and $D$ is the measure of
diffusivity. It is to be noted here that for $v = 0$ the system in Eq.~(\ref{langRB})
reduces to a pair of independent and identical Brownian searchers discussed above.
The two particles are independently searching for a static target located
at the origin and the process stops, when any of the two searchers find the target.
Depending on which particle reaches first, the winner is decided.

The unconditional FPTD of the $R_1-B_1$ system
described by Eq.~(\ref{langRB}) can be obtained using
the multiplicative and additive rules for probability of two independent events
constituting a Bernoulli trial:
\begin{enumerate}
\item $R_1$ reaches the target first,
\item $B_1$ reaches the target first.
\end{enumerate}
The reason we employ this method is in light of the difficulty arising in a direct
solution of the $R_1-B_1$ system, as the FPTD of
even a single RTP is difficult to obtain for a general $D$
\cite{malakar2018steady}. However, at long times an RTP with diffusion coefficient
$D$ behaves like a Brownian particle with diffusion coefficient $D+1/2$ 
\cite{malakar2018steady}. As a result, the long-time behavior of the
unconditional FPTD for the $R_1-B_1$ system described in (\ref{langRB}) representing
the above Bernoulli trial is
\begin{align}
F_{R_1B_1}(t) \stackrel{\text{large}~t}{\simeq} \frac{1}{t^2}.
\end{align}
This implies that the mean time to locate the target either by $R_1$ or $B_1$
diverges logarithmically (see Eq.~(\ref{mfpt})). However, unlike the case of identical
Brownian searchers discussed above, an RTP and a Brownian searcher starting at the
same initial location need not have equal search probabilities. For example,
for both the searchers starting at $c_1 = c_2 = 1$ and parameter values $v = 1,~
\gamma = 1$ and $D = 1$, numerical solutions of the Langevin equation (\ref{langRB})
lead to $p_{R_1} \approx 0.51$, implying that the RTP is the more efficient
searcher even when both $R_1$ and $B_1$ are starting at the same initial location.
In the extreme limit of $v\to \infty$ and $\gamma \to \infty$ such
that the ratio $v^2/2\gamma = D_{RTP}$ is fixed, the RTP behaves like a Brownian particle
with an effective diffusion coefficient of $D_{RTP} + D$, and thus, would naturally
have a higher search probability $p_{R_1}$ for $c_1 = c_2$.

The above discussion implies that amongst a pair of identical Brownian searchers
$B_1$ and $B_2$, the searcher starting closer to the target has a higher search
probability. On the other hand, for non-identical searchers such as an RTP and a
Brownian particle starting at the same initial locations, the RTP is the more
efficient searcher. In addition, once the system parameters are fixed,
the search probabilities are known \textit{a priori}. However, for purposes of
practical interest, methods for controlling the efficiency of searchers
is important, and restarts provide a viable mechanism to control the search
probabilities even when the system parameters remain fixed \cite{belan2018restart}.
The pressing question is, whether restarts can make a less efficient searcher
more efficient? In other words, if $p_1 \leq p_2$, can restarts result in
$p^R_1 \geq p^R_2$? Here and in what follows, the superscript $R$ denotes restarts.

\section{Competing searchers under restarts}
\begin{figure}
\includegraphics[width=0.5\textwidth]{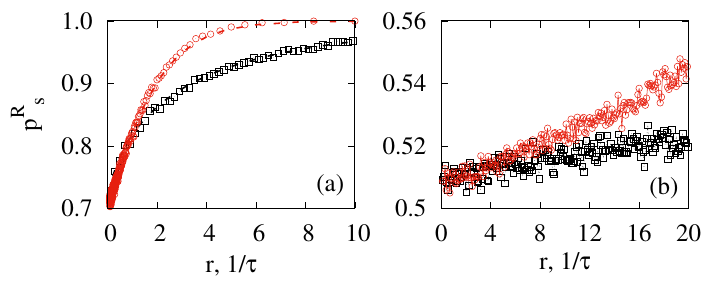}
\caption{(a) Success probability $p^R_s$ for Poisson ($\square$)
and sharp ($\circ$) restart protocols as function of restart rate ($r$ or $1/\tau$).
The dashed lines represent the formula in Eq.~(\ref{succ_prob}) while the
symbols are numerical estimates based on the solution of the Langevin
equation under restarts. Parameter values are $c_1 = 1,~c_2 = 2$, and $D = 1$.
(b) Success probability $p^R_s$ for Poisson ($\square$) and sharp ($\circ$)
restart protocols for a pair of competing RTP and Brownian searcher.
Parameter values are $c_1 = 1,~c_2 = 1,~v = 1,~\gamma = 1$, and $D = 1$.
Time step $dt = 4\times 10^{-3}$.}
\label{fig1}
\end{figure}
In order to study the effect of restarts on the search probabilities we consider
both Poisson restarts and sharp restarts. In Poisson restarts the two
searchers start all over again at a fixed rate while for sharp restarts
the motion renews after fixed time intervals. For the pair of Brownian searchers
$B_1-B_2$, let us define \textit{success} as the event
in which the search is completed by the searcher starting close to the
target, that is, $B_1$ since $c_1 < c_2$. On the other hand, for the $R_1-B_1$
we define \textit{success} as the event in which the search is completed by
the RTP. Now the success probability under Poisson
and sharp restarts reads \cite{belan2018restart,singh2023bernoulli}:
\begin{align}
\label{succ_prob}
p^R_s =\begin{cases}
\frac{\tilde{F}_{B_1}(r)}{\tilde{F}_{B_1B_2}(r)},~\text{Poisson restarts},
\vspace{0.25cm}\\
\frac{\int^\tau_0 dt~F_{B_1}(t)}{\int^\tau_0 dt~F_{B_1B_2}(t)},~\text{sharp restarts},
\end{cases}
\end{align}
where $\tilde{F}(r) = \int^\infty_0 dt~e^{-rt} F(t)$ denotes the Laplace
transform of $F(t)$. Similarly $\tilde{F}_1(r)$ is defined. In Eq.~(\ref{succ_prob})
$r$ denotes the rate of Poisson restart while $\tau$ defines the time-interval
of sharp restart, and $p^R_s$ is the success probability under restarts.
For the system of two Brownian searchers we find that the success probability $p^R_s$
is a monotonically increasing function of the restart rate ($r$ or $1/\tau$)
(see Fig.~\ref{fig1}(a)). For the $R_1-B_1$ system we resort to the numerical
solution of the Langevin equations in (\ref{langRB}) and find that $p^R_s$
is a monotonically increasing function of the restart rate for the case in which
the two searchers start from the same location (see Fig.~\ref{fig1}(b)).
The initial velocity of the RTP is $\pm v$ with probability $\frac{1}{2}$ each, and at every
restart event the RTP is relocated to its initial location with its velocity
remaining unchanged. In other words, only the position is subject to restarts.
The results in Fig.~\ref{fig1} imply that restarts tend to make
an already efficient searcher more efficient when the system is subject to
restarts, that is, $p^R_s \geq p_1$. Conversely, $1-p^R_s \leq p_2$.
Is this always the case or there could be a scenario in which a less efficient
searcher becomes more efficient at search under restarts?

In order to address this question
consider a pair of searchers $S_1$ and $S_2$ moving on the semi-infinite line
and searching for a target located at the origin. In the absence of any restarts,
both $S_1$ and $S_2$ have their respective search probabilities
$p_1$ and $p_2$ such that $p_1 > p_2$, that is, $S_1$ is the more
efficient searcher. At this point, the searchers $S_1$ or $S_2$
can either be a Brownian searcher or an RTP or something else. Then
if both the searchers are simultaneously subject to restarts,
then there is no selective advantage brought about
by restarts to any of the searchers. As a result, if a searcher was
earlier able to reach the target effectively, then the process of removing
the trajectories flying off to infinity will make it even more efficient,
that is, $p^R_1 > p_1$. However, this enhanced efficiency comes at the cost of
making it harder for $S_2$ to reach the target, that is, $p^R_2 < p_2$.
Summarily, we find $p^R_1 > p_1 > p_2 > p^R_2$ which implies that
\begin{align}
\label{pr1}
    p^R_1 > p^R_2.
\end{align}
It is to be noted here that the above inequality holds true for any
restart protocol, and not just Poisson or sharp restarts. Moreover, the inequality
in (\ref{pr1}) also implies that restarts tend to increase the difference
between search probabilities, that is, $p^R_1 - p^R_2 > p_1 - p_2$.

\section{\label{sec2a1}Subsystem restarts}
\begin{figure}
    \centering
    \includegraphics[width=0.45\textwidth]{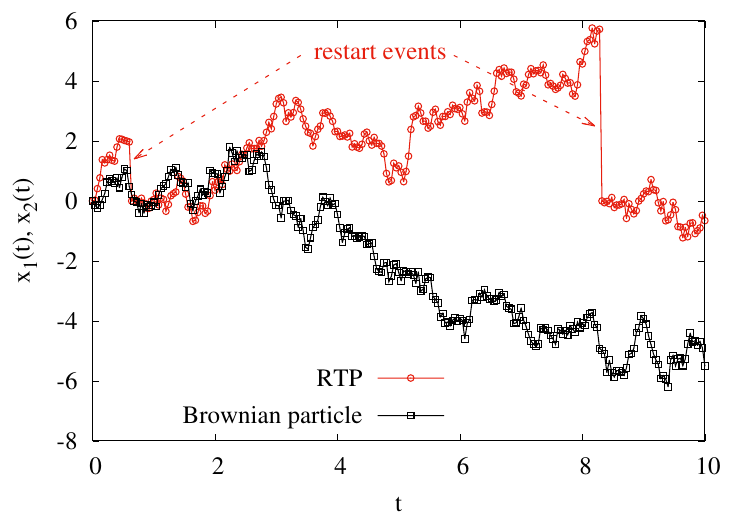}
    \caption{Typical trajectories of an RTP and a Brownian particle starting
    at the origin and the RTP is subject to restarts at fixed rates. A particular
    restart event is marked at around $t = 8$. Parameter
    values are $c_1=c_2 = 0$ and $v=1$, $\gamma=1$, $D = 1$. Time step $dt = 4\times 10^{-3}$.}
    \label{figtr}
\end{figure}
Whenever a dynamical system with more than one degree of freedom
is subject to restarts, a naturally
arising question is whether all the degrees are subject to restart simultaneously
or only a subset of them is restarted? For example, when
an RTP is subject to restarts, either only its position is restarted without affecting
the velocities (the case discussed here) or both position and velocity are restarted 
\cite{evans2018run}. However, if a dynamical system has multiple particles, then
subjecting a subset of particles to restarts leads to nontrivial implications on the
dynamical properties of the whole system. For example, restarting the phases
of a fraction of oscillators results in globally synchronizing the oscillators,
a concept termed as \textit{subsystem restarts} \cite{majumder2024kuramoto}. 
Applying subsystem restarts in the present context of competing
searchers would mean that only one of the searchers is subject to restart,
while the other evolves with its original dynamics unperturbed.
A typical representation of such a scenario is depicted in Fig.~\ref{figtr}
wherein a system of independent RTP and Brownian particle are moving on the line
and the RTP is subject to restarts at a constant rate.

\begin{figure}
\includegraphics[width=0.5\textwidth]{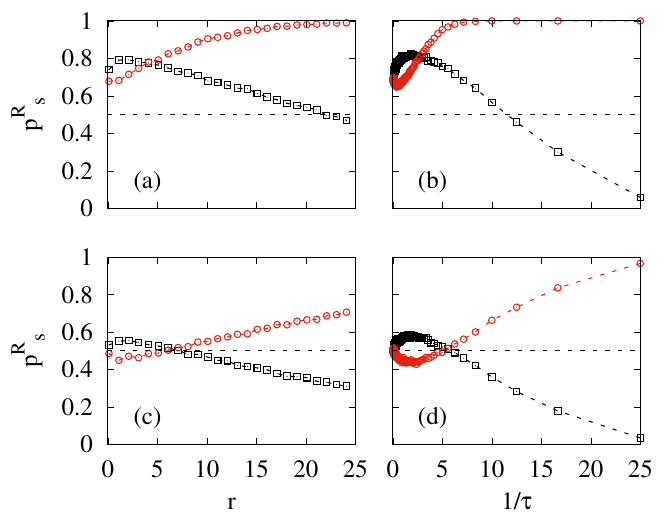}
\caption{Top panels: Success probability under subsystem restarts for a pair
of independent and identical Brownian searchers with (a) Poisson restarts
and (b) sharp restarts. For both restart protocols, black squares represent
the case when $B_1$ is restarted while red circles represent restarts of $B_2$.
The diffusion coefficient $D = 1$ and the searchers start respectively at
$c_1 = 1$ and $c_2 = 2$.
Bottom panels: The success probability under subsystem restarts for a system of RTP and Brownian
searchers under (c) Poisson restarts,
and (d) sharp restarts. For both restart protocols, black squares represent
the case when $R_1$ is restarted while red circles represent restarts of $B_1$.
Parameter values are $v=1$, $\gamma=1$, $D = 1$. The initial position of the two searchers is
$c_1=c_2 = 1$. Time step $dt = 4\times 10^{-3}$.
}
\label{fig2}
\end{figure}

In order to assess the effect of subsystem restarts on search probabilities
we consider: (i) a pair of independent and identical Brownian searchers $B_1$ and
$B_2$ starting respectively at $x_1(0) = c_1$ and $x_2(0) = c_2$ with $c_1 < c_2$;
and (ii) a pair of independent RTP and Brownian searchers $R_1$ and $B_1$
starting at the same position. Once again, we define \textit{success} when
(i) $B_1$ reaches the target and (ii) $R_1$ reaches the target.
For both scenarios we find in Fig.~\ref{fig2}
that the success probability $p^R_s$ exhibits a nonmonotonic dependence on the restart
rates $r$ or $1/\tau$ for both Poisson and sharp restart protocols respectively.
We see from Fig.~\ref{fig2}(a-b) that when $B_1$ is subjected
to restarts ($B_2$ evolves unperturbed), then for both Poisson
and sharp restart protocols the success probability $p^R_s$ exhibits a non-monotonic
dependence on the restart rate ($r$ or $1/\tau$), with the maximum occurring
in the region where the restart rate is small. Conversely, when restarts
take place sufficiently fast, then $p^R_s$ decreases monotonically.
This is because frequent restarts tend to confine the particle locally,
preventing it from reaching the target. In summary, restarts at a sufficiently
high rate make it less likely for $B_1$ to reach the target. Furthermore,
a sufficiently high rate could result in $p^R_s < 1/2$, implying that $B_2$
(searcher without restarts) is a more efficient searcher.
Conversely, if $B_2$ is subject to restarts, even though it has a
marginal advantage for small restart rates (see minimum
of red circles in Fig.~\ref{fig2}(b)), $p^R_s > 1/2$ for all the restart rates,
that is, $B_1$ still remains the efficient searcher. Similarly,
for the $R_1-B_1$ system of RTP and Brownian searchers, the otherwise
efficient RTP can become less efficient when subject to restarts
(see Fig.~\ref{fig2}(c-d)). Interestingly, the existence of the extremum of $p^R_s$
seems to be independent of the identity of individual searchers and the
details of the restart protocol. Moreover, when the more efficient searcher
$S_1$ is restarted, then $p^R_s$ exhibits a maximum, approaching zero for high
restart rates for both Poisson and sharp restart protocols. Conversely,
when the less efficient searcher $S_2$ is subject to restarts, then
the success probability $p^R_s$ exhibits a minimum and approaches unity for
large restart rates.
Does this mean that the observed extremum of $p^R_s$ is a generic property
of subsystem restarts?

\begin{figure}
\includegraphics[width=0.5\textwidth]{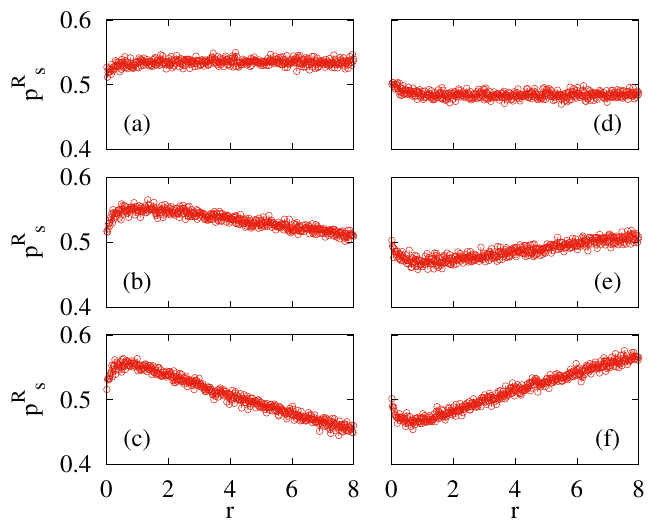}
\caption{Success probability as a function of rate $r$ for the $R_1-B_1$
system with the RTP under restarts for (a) $\alpha = 1/2$,
(b) $\alpha = 3/2$, (c) $\alpha = 5/2$; and the Brownian particle under restarts
for (d) $\alpha = 1/2$, (e) $\alpha = 3/2$, (f) $\alpha = 5/2$. Parameter values
are $v=1$, $\gamma=1$, $D=1$ and the searchers initially start at $c_1=c_2 = 1$.
Time step $dt = 4\times 10^{-3}$.}
\label{fig4}
\end{figure}
In order to answer this question, consider the $R_1-B_1$ system subject
to subsystem restarts wherein the distribution of restart times $R$ is
a power law distribution:
\begin{align}
\label{pwrlw}
P(R) = \frac{\alpha r}{(1+rR)^{1+\alpha}}
\end{align}
with $r,~\alpha > 0$. For $\alpha > 2$ both $\langle R \rangle$ and
$\langle R^2 \rangle$ exists, while for $\alpha > 1$ only $\langle R \rangle$
is finite. For $\alpha \in (0,1)$ the distribution $P(R)$ does not
possess any moments. We study this system by numerically solving
the Langevin equations (\ref{langRB}) and define \textit{success} as
the event in which the RTP reaches the target. We find from Fig.~\ref{fig4}
that  the success probability $p^R_s$ exhibits a nonmonotonic
dependence on the restart rate $r$ for both cases when either the RTP is restarted
(see Fig.~\ref{fig4}(a-c)) or when the Brownian
searcher is subject to restarts (see Fig.~\ref{fig4}(d-f)). Moreover,
even though $p^R_s$ in Fig.~\ref{fig4}(a-b) seems to saturate to a
fixed value, numerical calculations for large $r$ indicate that indeed
$p^R_s$ exhibits a decay for the case when the RTP is subject to
restarts while it increases with increasing $r$ when the Brownian
searcher is restarted.

The results in Figs.~\ref{fig2} and \ref{fig4} suggest that the nonmonotonic
dependence of the success probability $p^R_s$ on the restart rate
seems to be a generic feature of subsystem restarts. To address this
in general, consider a pair of searchers $S_1$ and $S_2$ moving
on the real line and searching for a target at the origin. Under
subsystem restarts, only one of the searchers $S_i$
is subject to restarts while the other $S_j$ evolves unperturbed, with
$i,j~\in \{1,2\}$ and $i \neq j$. Then the fact that $S_i$ is subject to
restarts implies that its trajectories meandering off to infinity
are cut short due to restarts at some point of time. As a result,
the introduction of restarts will improve its search efficiency. This is
particularly important for small restart rates $R$, when
introduction of restarts tend to improve the search efficiency by
removing trajectories meandering off to infinity. In other
words, if $S_i$ is restarted at a rate $R$ then
\begin{align}
\label{ineq2}
\lim_{R \to 0} p^R_i \geq p_i.
\end{align}
On the other hand, if the restart rate $R$ is very large,
then $S_i$ will be localized to its initial location, for all practical
purposes, and as a result, shall never reach the target, which means
\begin{align}
\label{ineq3}
\lim_{R \to \infty} p^R_i = 0.
\end{align}
From the inequalities (\ref{ineq2}) and (\ref{ineq3}) it follows
that if $S_1$ is subject to restarts while $S_2$ evolves unperturbed,
then $p^R_1$ will exhibit a local maximum. Conversely, when $S_2$ is subject
to restarts with $S_1$ evolving unperturbed then $p_1$ will exhibit a minimum.
It is to be noted here that we have not assumed any specific details about
the distribution of restart times or the identity of individual searchers.
The only conditions on the inequalities (\ref{ineq2}) and (\ref{ineq3}) are
that the searchers are moving independently on the semi-infinite
line and $R$ is a quantity with dimension of inverse time. Depending on
the details of the restart protocol, $R$ can either be $r$ (Poisson restarts
or power-law restarts) or $1/\tau$ (sharp restarts).

\section{First passage times under subsystem restarts}
\begin{figure}
    \centering
    \includegraphics[width=0.45\textwidth]{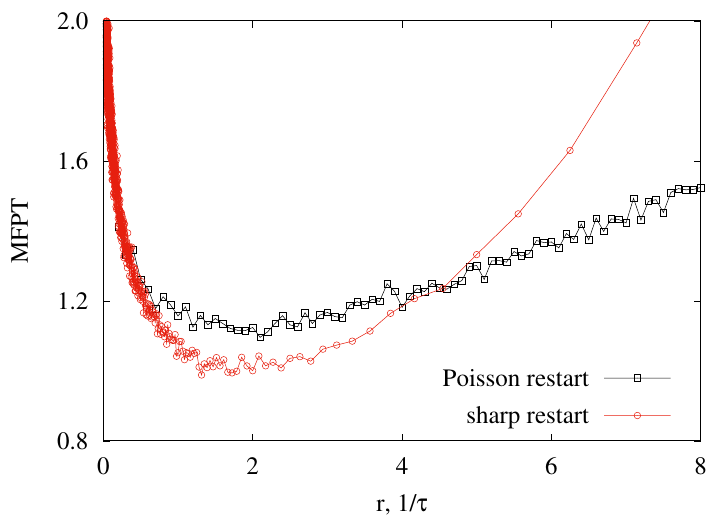}
    \caption{MFPT under subsystem restarts for the $R_1-B_1$ system in which
    the RTP $R_1$ is subject to restarts while the Brownian searcher $B_1$
    evolves unperturbed. Parameter values are $v=1$, $\gamma=1$, $D=1$
    and the searchers initially start at $c_1,~c_2 = 1$. Time step $dt = 4\times 10^{-3}$.}
    \label{figMFPT}
\end{figure}
So far we have extensively studied the properties of success probability
for a pair of competing searchers under restarts. However, the concept
of first passage times is natural to search processes and provides a complementary
view point towards our understanding of Bernoulli trials. While the theory
of first passage times is well developed for Bernoulli trials under
restarts \cite{belan2018restart,singh2023bernoulli}, the concept
of subsystem restarts is still in its infancy, let alone its first passage
properties. In order to understand the behavior of first passage times
under subsystem restarts, let us study the earlier example of the $R_1-B_1$
system wherein the RTP is subject to restarts while the Brownian searcher
evolves unperturbed. As an RTP with a diffusion coefficient $D$ behaves
like a Brownian particle with an enhanced diffusion coefficient $D + 1/2$
at long times, this implies that its FPTD
$F_{R_1}(t) \stackrel{\text{large}~t}{\simeq} \frac{1}{t^{3/2}}$
\cite{malakar2018steady}. Now, with
the introduction of restarts (Poisson or sharp), the long time behavior
of the FPTD is modified so as to exhibit an exponential decay, that is,
$F^R_{R_1}(t) \stackrel{\text{large}~t}{\simeq} e^{-t}$~\cite{evans2011diffusion,
singh2022capture}. As a result, the corresponding survival probability $q_{R_1}(t)$
shall also exhibit an exponential decay. Now the Bernoulli trial $R_1-B_1$
subject to subsystem restarts stops if any of the following events takes place:
\begin{enumerate}
    \item $R_1$ subject to restarts reaches the target, or
    \item $B_1$ reaches the target.
\end{enumerate}
The unconditional FPTD for the $R_1-B_1$ Bernoulli trial under subsystem
restarts reads:
\begin{align}
\label{fr0b0}
    F^R_{R_1B_1}(t) = F^R_{R_1}(t)q_{B_1}(t) + q^R_{R_1}(t)F_{B_1}(t)
    \stackrel{\text{large}~t}{\simeq} \frac{e^{-t}}{\sqrt{t}}.
\end{align}
It is to be noted here that we do not furnish any information regarding
the dependence of $F^R_{R_1B_1}(t)$ on the restart rate ($r$ or $1/\tau$), which
affects both the FPTD and the success probability in a nontrivial manner.
Notwithstanding, the generic structure of Eq.~(\ref{fr0b0}) does imply
towards the existence of moments of the FPTD. For example, the mean first
passage time (MFPT) $\langle t \rangle = \int^\infty_0 dt~tF^R_{R_1B_1}(t)\simeq \int dt~\sqrt{t}e^{-t}$ is finite. We study the MFPT by numerical solution
of the Langevin equation (\ref{langRB}) and find that the MFPT exhibits a
non-monotonic dependence on the restart rate (see Fig.~\ref{figMFPT}),
a characteristic signature of MFPT under restarts.

\section{Conclusions}
We study competing searchers in one spatial dimension and searching for
a target located at the origin. Starting with a pair of independent and
identical Brownian searchers at different initial positions we find that
the application of restarts tend to make an already efficient searcher more
efficient. In other words, the search probability for a searcher starting
close to the target increases with the introduction of restarts. We demonstrate
this to be true for a pair of non-identical searchers, namely, an RTP and
a Brownian searcher and show that restarts, in general, make an already
efficient searcher more efficient. However, subjecting both the independent
searchers to restarts simultaneously, in addition to preserving the order
of search probabilities, increases their difference. In order to alter
this order, we introduce subsystem restarts in which only one of the
searchers is subject to restarts while the other evolves in an unperturbed
manner. We show that in the limit of small restart rates $R$,
the searcher subject to restarts experiences a marginal increase in its
search probability. Conversely, in the limit of large restart
rates the searcher becomes stalled (for all practical purposes) resulting in
its search probability approaching zero.

In this work we have focused our attention on probabilities of successful
completions of a Bernoulli trial. For example, for the case of an RTP
and a Brownian searcher the information about search probabilities and its
dependence on restarts provides us with a picture complementary to the analysis
of first passage times. Even though we have a few general results about the
search probabilities for a pair of competing searchers,
it would be interesting to study subsystem restarts for $N > 2$ searchers
\cite{grebenkov2020single}.
We have also seen above that the MFPT under subsystem restarts exhibits a
nonmonotonic dependence on the restart rate $R$ for both Poisson
and sharp restarts. This implies that the optimal restart rates can also
be defined for subsystem restarts and it might exhibit a dynamical phase
transition reported earlier for Bernoulli trials \cite{singh2023bernoulli}.
We take up these and related questions in future works.

\begin{acknowledgments}
RM and TS acknowledge financial support by the German Science Foundation (DFG, Grant number ME 1535/12-1). TS is also supported by the Alliance of International Science Organizations (Project No. ANSO-CR-PP-2022-05) and by the Alexander von Humboldt Foundation.
\end{acknowledgments}



\appendix

\bibliography{aipsamp.bib}

\end{document}